\documentstyle[12pt]{article}

\renewcommand{\thefootnote}{\fnsymbol{footnote}}

\newcommand{\newsection}{    % Numeration of eqs. is automatic
\setcounter{equation}{0}
\section}

\def\LB{\left (}
\def\RB{\right )}
\def\K{\tilde K}
\def\W{\tilde W}
\def \ov {\over }
\def\T{{\rm T}}

\newcommand{\non}{\nonumber \\*}

\newcommand{\eq}[1]{Eq.~(\ref{#1})}

\def\bea{\begin{eqnarray}}
\def\eea{\end{eqnarray}}

\def\be{\begin{equation}}
\def\ee{\end{equation}}

\def\td{\tilde}

\def\det{\hbox{det}}

\def\la{\label}

\def\np {{  Nucl. Phys. }}

\def \pr  {{ Phys. Rev. }}
\def \bi{\bibitem}

\begin{document}
\begin{titlepage}
\begin{flushright}
ITP-SB-99-01\\
\end{flushright}
\vspace{.5cm}

\begin{center}
{\LARGE  Are M-atrix theory and Maldacena's conjecture related? }\\
\vspace{1.1cm}
{\large Iouri Chepelev\footnote{
E-mail: chepelev@insti.physics.sunysb.edu }}\\
\vspace{18pt}
{\it Institute for Theoretical Physics}

{\it SUNY at Stony Brook}

{\it  NY11794-3840, USA}
\end{center}
\vskip 0.6 cm

\begin{abstract}
We give arguments in the support of a relation between
M-atrix theory and Maldacena's conjecture. M-atrix theory
conjecture implies the equivalence of 11-D light-cone supergravity and
strongly-coupled (0+1)-D SYM. Maldacena's  SUGRA/SYM duality conjecture implies, in the one dimensional SYM case, the equivalence
between strongly-coupled (0+1)-D SYM and 11-D supergravity compactified on 
a spatial circle in the
formal Seiberg-Sen limit.  Using the classical equivalence
between 11-D supergravity on a  light-like circle
and on  a spatial circle in the formal Seiberg-Sen limit,  
we argue that in the (0+1)-D SYM case, the
 large-N M-atrix theory in the  supergravity regime is equivalent to SUGRA/SYM duality.

\vspace{2cm}

{\small {\it PACS}: 04.50.+h; 04.65.+e; 11.25.-w; 11.30.Pb

}

\end{abstract} 

\end{titlepage}
\setcounter{page}{1}
\renewcommand{\thefootnote}{\arabic{footnote}}
\setcounter{footnote}{0}

\newsection{Introduction}
M-atrix theory \cite{bfss,suss} and Maldacena's conjecture \cite{malda} are 
similar in the sense that both are
about the equivalence between
M/string theories (i.e. theories in the bulk of space-time) and  brane
world-volume theories. But does this  similarity 
go beyond a mere analogy? Maldacena's conjecture relates M/string
theory on a curved background to SYM in the flat space, whereas M-atrix
theory relates M-theory on a transverse torus to SYM on a dual torus.\footnote{In what follows we will consider  M-atrix theory in  flat backgrounds. M-atrix theory
in  curved backgrounds is still very poorly understood \cite{doug}.}
For example, consider finite-N M-atrix theory on a 3-torus. If $V_3$ is
the volume of M-theory transverse 3-torus and $R$--the radius of
light-like circle, then the corresponding D=4
SYM is defined on a dual 3-torus of volume $\td V_3 = (R M_P^3)^{-3} V_3^{-1}$.
Dual 3-torus decompactifies in the limit $V_3\rightarrow 0$ , but 
$g_{\rm YM}^2 = (M_P^3 V_3)^{-1}$ also blows up in this limit. In the 
large-N limit one has to take $R\rightarrow \infty$. Thus we have
$g_{\rm YM}^2 \gg 1$ and  $N\gg 1$, and this  limit is different from the one
involved in ${\rm AdS}_5/{\rm CFT}_4$ correspondence. 

\begin{picture}(300,180)(-30,0)
%%left big oval%%
\put(80,90){\oval(40,120)}
\put(77,155){${\cal M}_1$}
%%small left upper oval%%
\put(80,105){\oval(20,44)}
%%small left lower oval%%
\put(80,75){\oval(20,44)}
%%right big oval%%
\put(250,90){\oval(40,120)}
\put(247,155){${\cal M}_2$}
%%small right upper oval%%
\put(250,105){\oval(20,44)}
%%small right lower oval%%
\put(250,75){\oval(20,44)}
%%matrix map first  from the top%%%
\put(80,127){\vector(1,0){50}}
\put(130,127){\line(1,0){120}}
%%matrix map second  from the top%%%
\put(80,83){\vector(1,0){50}}
\put(130,83){\line(1,0){120}}
%%maldacena map first line%%
\put(80,97){\vector(1,0){120}}
\put(200,97){\line(1,0){50}}
%%maldacena map second line%%
\put(80,53){\vector(1,0){120}}
\put(200,53){\line(1,0){50}}
%%%%
\put(170,60){${\cal F}_{\rm Maldacena}$}
\put(115,134){${\cal F}_{\rm M-atrix}$}
\put(60,10){Figure 1: M-atrix and Maldacena maps}
\end{picture}

These formal arguments do not,
however, imply that the two conjectures are in conflict with each other,
but rather suggest the following possibility. Imagine two moduli spaces:
${\cal M}_1$ of M-theory and ${\cal M}_2$ of SYM (coordinates of these
 moduli 
spaces are: the dimension $d$ of SYM, the rank $N$ of gauge group, 
the type of gauge theory base space, the type of space on which
M-theory is compactified and so on). One can think of the two 
conjectures as two one-to-one maps
${\cal F}_{\rm M-atrix}$ and ${\cal F}_{\rm Maldacena}$
between certain regions of ${\cal M}_1$
and $ {\cal M}_2$. It is clear that if these mappings map two non-overlapping
 regions in ${\cal M}_1$ into two non-overlapping regions in ${\cal M}_2$, 
it does not mean that they contradict each other (see Figure 1). A natural 
question to ask
is whether 
the domains of these 
two mappings intersect, and if so, are the two mappings identical in the 
overlap? The motivation for
writing this paper was to answer this question.\footnote{For some other discussions of M-atrix/Maldacena relation,
see refs.~[5--10].}

In this paper we will consider (0+1)-D SYM case.\footnote{Supersymmetric
matrix quantum mechanics was first studied in ref.\cite{halpern}.}   
One dimensional SYM is simplest in the sense that 
no compactification is involved. The
purpose of this
paper is to establish a relation between large-$N$, uncompactified
M-atrix theory in the supergravity  regime and SUGRA/SYM duality (large-$N$
version of Maldacena's conjecture)
in  this one dimensional case.  

The paper is organized as follows. In Sec.2
we give a brief review M-atrix theory. In Sec.3 we discuss some
aspects of SUGRA/SYM duality
 and propose a new interpretation of SUGRA/(0+1)-D SYM duality.
In Sec.4 we argue that M-atrix theory and SUGRA/(0+1)-D SYM duality
are related by a classical equivalence between 11-D supergravities on
a light-like and a spatial circles. 

\newsection{M-atrix theory}
The finite-$N$ version of M-atrix theory conjecture \cite{suss}, as clarified by Sen and Seiberg \cite{sen,seiberg}, states that
discrete light-cone quantization (DLCQ) of M-theory (with the Planck scale $M_P$ and a transverse length scale $x$)
 compactified on a light-like
circle ($x^- \equiv x^- + 2 \pi R$) in
the sector with the total longitudinal momentum $p_-= { N \ov R}$ is
equivalent to 
$\td M$-theory (with the Planck scale $\td M_P$ and a transverse length scale
$\td x$) compactified on a spatial circle $(x_{11}\equiv x_{11} + 2\pi \td R$) 
in the sector with the total Kaluza-Klein
momentum along the $x_{11}$ direction $p_{11}={N\ov \td R}$ in the limit
$\td R \rightarrow 0$ , $\td M_P\rightarrow \infty$ , 
with the fixed
\be
\td R \td M^2_P = R M^2_P\ , \ \ \  \ \ \  
\td x \td M_P= x M_P   \  .        
\la{para}
\ee
The large-$N$ version of M-atrix theory conjecture \cite{bfss} is a related 
statement about the {\it uncompactified} 11-D light-cone M-theory. 
It is {\it assumed} that the large-N version of M-atrix theory is the
$R, N\rightarrow \infty$ (with $N/R={\rm fixed}$) limit of finite-$N$ theory.

Compelling arguments for the validity of finite-$N$ M-atrix theory
 conjecture were given by Seiberg and Sen \cite{seiberg,sen}.\footnote{
We follow a very nice review of these arguments  given in Sen's review article 
\cite{sen2}.} Let ${\cal T}$ be a field theory (with a mass scale and a set of
dimensionless parameters $\{ y_i \}$) formulated on a flat background
(or on a background  of the form $R^{1,1}\times {\cal M}$ which 
has $SO(1,1)$ isometry). 
Let
${\cal H}_N^{DLCQ}(M, R, \{ y_i\} )$ be  the N-body  DLCQ quantum 
mechanical hamiltonian
describing the dynamics of the theory ${\cal T}$ compactified   
on a light-like circle ($x^-\equiv x^- +2\pi R$) in the sector with the
total longitudinal momentum $P_- = {N\ov R}$.
The dynamics of the theory ${\cal T}$ compactified
on a small spatial circle ($x\equiv x+2\pi \td R$) in the sector with 
$N$ units of total momentum along the $x$ direction is also expected to be described by a non-relativistic $N$-body 
hamiltonian.\footnote{The argument goes as follows: $p_{11}={N\ov \td R}$, $H=\sqrt{p_{11}^2+p_{\perp}^2}=p_{11}
+{p_{\perp}^2\ov 2 p_{11}}+\cdots $ and in an appropriate limit when $\td R\rightarrow 0$ one is left with a non-relativistic hamiltonian.}
We denote it by ${\cal H}_N^{KK}(\td M, \td R, \{ y_i\} )$.
 Now, thinking of the light-like circle as an almost light-like circle and
boosting it, one can show that 
\be
{\cal H}_N^{DLCQ}(M, R, \{ y_i\} )=\lim_{\td R \rightarrow 0} {\cal H}_N^{KK}(\td M=M\sqrt{{R\ov \td R}}, \td R, \{ y_i\} ) \ .
\la{relation}
\ee
\eq{relation} follows from the following  kinematical relation between 
two coordinate frames
related by a boost $\beta$ :
\be
{\partial \ov \partial x'^+}=e^{\beta} \LB {\partial \ov \partial t}+{\partial
\ov \partial x}\RB  \ ,
\la{boost}
\ee
and the assumption that the quantum operators which generate $i(\partial /\partial x'^+)$, $i(\partial /\partial t)$ and $i(\partial /\partial x)$ are 
${\cal H}_N^{DLCQ}$, ${\cal H}_N^{KK}+N/\td R$ and $-N/\td R$, respectively.

\newsection{SUGRA/SYM duality}
The essence of SUGRA/SYM duality is the following:
surround a D-brane source in supergravity by a sphere of radius
$\sqrt{\alpha'}$. Supergravity is a low energy limit of string theory and
hence, a priori, it is valid only outside of the sphere. It is {\it claimed}
that if the curvature and effective string coupling of the supergravity 
solution 
are small, supergravity is valid also inside of the sphere. 
It should be
noted that this idea does not give a precise recipe for
the SUGRA/SYM correspondence. The rules of the game should be
specified for a particular supergravity background, etc. For example, in 
the AdS/CFT
case, such a recipe was given in refs.\cite{gkp,wit}. In this case, the 
correspondence
is between the fluctuations of supergravity fields on the {\it curved} AdS 
background and
SYM operators.
We are interested in SUGRA/(0+1)-D SYM duality which was discussed in
ref.\cite{IMSY}. In the latter paper it was argued that the motion of
a D0-brane probe carrying one unit of RR-charge  in the curved near-horizon 
geometry  
of a source D0-brane of charge $N\gg 1$ is dual to a process in (0+1)-D SYM 
with
the gauge group $U(N+1)$ broken to $U(N)\times U(1)$ by a Higgs vev.
Motivated by AdS/CFT correspondence, one may argue that (0+1)-D SYM
is dual to supergravity on a curved near-horizon D0-brane background.   
It seems that this is a widely accepted  point of view.\footnote{
For a discussion of ${\rm AdS}_2/{\rm SYM}_{0+1}$ duality see ref.\cite{domain}. We will 
argue that ${\rm AdS}_2/{\rm SYM}_{0+1}$ duality  conjecture of ref.\cite{domain} can 
be  correct only approximately.} Let {\bf A} be a
proponent of this point of view and {\bf B}--a proponent  of the following
interpretation (which is our interpretation):
 SUGRA/(0+1)-D SYM duality  is a relation
between the scattering  of 11-D supergravity states in a  
{\it flat} background and the corresponding SYM processes. The meaning
of our interpretation is best illustrated by the following
discussion between {\bf A} and {\bf B}. 

{\bf A}: Taking Maldacena's low energy limit in the D0-brane metric, one 
    obtains the metric of curved near-horizon geometry of D0-brane. Thus,
    (0+1)-D SYM is dual to supergravity on the curved space. For example,
    consider the scattering of three D0-branes. Let the first two of these 
    D0-branes
    carry one
    unit of RR-charge each and the third one carries $N\gg 1$ units of 
    RR-charge. One can  think of the scattering of the first two
    D0-branes as taking place
    in the curved near-horizon geometry of the third D0-brane. One
    should for instance use a curved-space propagator to compute scattering
    amplitudes. This curved-space supergravity picture is dual to (0+1)-D
    SYM  with the gauge group $U(N+2)\rightarrow U(N)\times U(1)\times U(1)$.
\newline
{\bf B}: I agree  that  in the case of $U(N+2)\rightarrow U(N)\times U(1)\times 
   U(1)$ your curved-space picture works. But I have problem with
   the other cases. Suppose that the gauge group $U(N)$ is broken to 
   $U(N_1)\times U(N_2)\times \cdots \times U(N_k)$ with $N_i \sim N$.
   In this case I cannot talk about scattering as taking place in 
   a definite background since the back-reaction effects are not negligible. \newline
{\bf A}:  Well, I think SUGRA/(0+1)-D SYM duality in this case should be 
   interpreted as
   a duality between the excitations of supergravity fields on
   the 'near-horizon' multi-center D0-brane static background and SYM 
   processes.
   In ref.\cite{msw} the centers of multi-center static solution were 
   promoted to dynamical variables. By the way, an analogous point
   of view, in the context of ${\rm AdS}_5/{\rm CFT}_4$ correspondence, was first
   advocated in Maldacena's paper \cite{malda}.\newline
{\bf B}: You forgot to mention one important point about ${\rm AdS}_5/
{\rm CFT}_4$
   correspondence on the Coulomb branch. ${\rm SYM}_4$ has moduli space
   of supersymmetric vacua which is identical to the parameter space of
   the general multi-center D3-brane solution. But we know that (0+1)-D SYM 
   does not have moduli space of vacua  because
   symmetry breaking scalar expectation values are not frozen variables \cite{banks}.
    Therefore, one cannot associate the parameter space of multi-center D0-brane 
   solution with the moduli space of vacua of (0+1)-D SYM.\newline
   {\bf A}: But may be instead of associating this parameter space with the moduli
   space of vacua (which does not exist in SUSY matrix quantum mechanics), one
   should associate it with the space of
   slow variables in a Born-Oppenheimer approximation. It was argued 
   in ref.\cite{banks} that Born-Oppenheimer approximation is justified in
   the large-N limit.\newline 
{\bf B}: I have little to say about it. Instead of trying to resolve
problems inherent to the curved-space picture, I want to propose  the 
following simple flat-space picture. 
   Let
   $A_n^{SUGRA}(\td R,\cdots )$ be a $n$-graviton scattering amplitude in the flat 11-D 
   supergravity compactified on a spatial circle of radius $\td R$. 
   Let $A_n^{SQM}$ be a corresponding SUSY quantum mechanical  amplitude. A simple
   recipe for the SUGRA/(0+1)-D SYM reads 
\be
   A_n^{SQM}=\lim_{\td R\rightarrow 0}A_n^{SUGRA}(\td R, \cdots ) \ .
\la{recipe}
\ee
   The curved-space picture is approximately correct and useful in some
particular cases like the case of  broken $U(N+2)\rightarrow U(N)\times
U(1)\times U(1)$, but  the flat-space picture is more general and contains 
the curved-space picture as a special case. Note  that
by  the flat space picture, I actually mean the procedure of  calculating scattering amplitudes 
in the
supergravity on a flat background and  taking Seiberg-Sen limit in 
the resulting amplitudes. In the flat-space picture context, the curved-space
picture emerges effectively in some cases as a result of taking $\td R\rightarrow 0$ limit. The  supergravity description that follows from \eq{recipe} should
be tested for its validity in each particular scattering problem. The 
analysis of ref.\cite{IMSY} shows that supergravity description
breaks down at long distances $r>N^{1\over 3}l_p$. I think the latter 
inequality should be replaced by a more general one in the case of  
$U(N)\rightarrow U(N_1)\times  \cdots \times U(N_k)$.
 \newline
{\bf A}:Did not Maldacena use the same arguments in his analysis
\cite{malda}? Namely, he started with an asymptotically flat D-brane
metric, took a low energy limit $\alpha'\rightarrow 0$ in it and obtained
an asymptotically curved metric. \newline
{\bf B}: The relation \eq{recipe} is a generalization of 
Maldacena's idea of taking low energy limit in the D-brane supergravity
solution to the cases when there is no well-defined background in which
 scattering takes place. Moreover, I am emphasizing flatness of supergravity 
because the
justification\footnote{The discussion of this justification will be given
below.} for taking $\td R \rightarrow 0$ limit in the supergravity
amplitudes involves a classical equivalence between supergravities on 
a light-like and
a space-like circles, and 
the latter equivalence requires the formulation
of supergravity on a flat background.

\newsection{M-atrix theory--SUGRA/SYM duality relation}

We see that in the M-atrix
theory context (0+1)-D SYM in the large-$N$ limit is equivalent to 
11-D light-cone 
supergravity, and
in the context of SUGRA/SYM duality it is equivalent to
11-D supergravity on a spatial circle in the formal Seiberg-Sen limit.
Assuming that M-atrix theory and SUGRA/SYM duality are related, this 
suggests that there should exist a relation between light-cone 11-D
supergravity and 11-D supergravity on a spatial circle in the formal
Seiberg-Sen limit (see Figure 2). Indeed,  
it was noticed
in refs.\cite{bbpt,CT}
that the scattering potentials calculated in these two supergravities
agree. 
It is also known that infinitely boosting the D0-brane supergravity solution is equivalent
to taking a near-horizon limit (Seiberg-Sen limit) \cite{bbpt}. 
The authors of refs.\cite{hyun,HK} used this fact to argue that SUGRA/SYM duality 
and  M-atrix theory are equivalent. Their arguments can be summarized as 
follows.
``Motivated by the finite-N version of M-atrix theory, let us up-lift  
the D0-brane 
SUGRA metric to 11-D and then compactify it on $x^-$. As a result, we find
that the harmonic function $H_0$ is replaced by $H_0 -1$. Now rewrite
this  metric in terms of new symbols ($\td r$, $\td g_s$, etc.) that
are related to the original ones ($r$, $g_s$, etc.) by  Seiberg-Sen's 
relations \eq{para}. Let us appropriately rescale 
(using Seiberg-Sen's relations)
the resulting 
 metric written in terms of new symbols.  We find that this rescaled metric is
precisely that of the D0-brane in the the near-horizon limit!
So, we see the validity of Seiberg-Sen's prescription when
applied to the background geometry of supergravity associated with the
DLCQ of M-theory. We also see that (0+1)-D SYM is equivalent to
fluctuations\footnote{We think these authors implicitly assumed this
when they talked about the background geometry etc.} of SUGRA fields on the 
curved near-horizon D0-brane background.
Since we know that the latter statement is precisely that of SUGRA/SYM 
duality,  we seem to have shown that SUGRA/SYM duality follows 
from M-atrix theory.''

There are two objections against their line of reasoning. First, the  
meaning of taking Seiberg-Sen limit in the expression for the 
metric is not clear. In Seiberg-Sen's derivation of M-atrix theory, this
limit was taken in the {\it full} M-theory and not just in supergravity. 
Second, 
the discussion of the authors of refs.\cite{hyun,HK} seems to imply
 that two M-theories formulated on the {\it curved} backgrounds are
related via Seiberg's boost, but
\begin{picture}(500,230)(0,-10)
\put(60,110){\vector(0,-1){40}}
\put(60,140){\vector(0,1){40}}
\put(31,126){$R$,$N\rightarrow\infty$ and}
\put(28,115){ low energy limit}
\put(380,110){\vector(0,-1){40}}
\put(380,140){\vector(0,1){40}}
\put(347,126){SUGRA/SYM}
\put(360,115){duality}
\put(60,50){\oval(70,40)}
\put(45,56){DLCQ}
\put(38,38){M-theory}
\put(380,50){\oval(70,40)}
\put(356,54){$\td M$-theory}
\put(364,37){(SYM)}
\put(60,200){\oval(70,40)}
\put(380,200){\oval(70,40)}
\put(38,206){light-cone}
\put(42,188){SUGRA}
\put(354,206){SUGRA on}
\put(356,188){spatial $S^1$}
\put(190,50){Seiberg-Sen}
\put(185,50){\vector(-1,0){89}}
\put(260,50){\vector(1,0){86}}
\put(170,200){classical equivalence}
\put(168,200){\vector(-1,0){72}}
\put(275,200){\vector(1,0){67}}
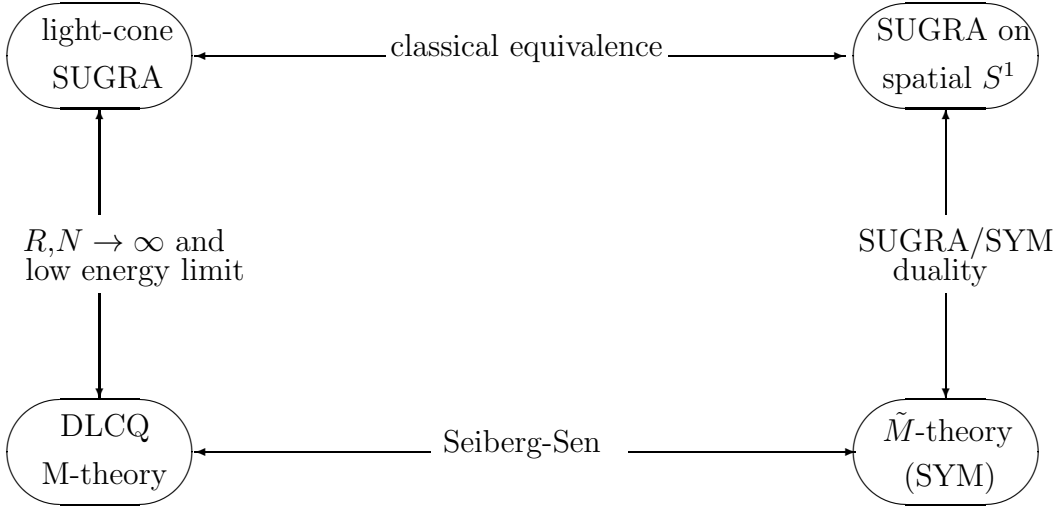
\put(130,10){Figure 2: Rectangle of relations}  
\end{picture}
 the backgrounds they considered 
do not have SO(1,1) isometry. But is  not it  the case that in the 
Seiberg-Sen's
derivation of Matrix theory one breaks Lorentz invariance by compactification
on a circle? It is so, but the underlying theory is supposed to be an
 SO(1,1) invariant theory formulated on a background of the form $R^{1,1}\times {\cal M}$. The backgrounds considered in refs.\cite{hyun,HK} are not
SO(1,1) isometric  before the compactification. 

We resolve the first problem by clarifying the meaning of taking Seiberg-Sen
limit in  supergravity expressions. Its meaning is the classical 
equivalence of 11-D supergravities on  light-like and spatial circles.
The second problem is resolved by interpreting SUGRA/(0+1)-D SYM duality
in a  way different from what they did. We stated earlier our interpretation
of this duality in the one-dimensional case: ...a relation
between the scattering  of supergravity states in a  
 flat background and the corresponding SYM processes.

In order to show that M-atrix theory and SUGRA/SYM duality are related one
has to use
 the ``rectangle of relations'' (see Figure 2) and 
the classical equivalence of two supergravities.
 Let us prove the classical equivalence of two supergravities.
Let us denote by $H_N^{l.c.}(M, R, \{ y_i\})$ and $H_N^{s.l.}(\td M, \td R, \{ y_i\} )$  classical hamiltonians of a  theory compactified on  light-like
and spatial circles in the
sectors with $N$ units of longitudinal momentum. 
In the hamiltonian formulation we associate them with the vector fields
$(\partial /\partial x^+)$ and $(\partial / \partial t)$.
The analog of \eq{relation} in this classical case reads
\be
H_N^{l.c.}(M, R, \{ y_i\} )=\lim_{\td R \rightarrow 0} H_N^{s.l.}(\td M=M\sqrt{{R\ov \td R}}, 
\td R, \{ y_i\} ) \ .
\la{relation2}
\ee
Note that the statements expressed by \eq{relation} and \eq{relation2} are
very different. The latter states the equivalence of two {\it classical}
field theory hamiltonians and the former states the equivalence of two 
{\it quantum
mechanical} $N$-body hamiltonians. It is the exploitation of this 
{\it difference}
that enables us to relate the two conjectures.
Applying \eq{relation2} to 11-D supergravity formulated in a flat
background, one establishes the classical equivalence between the two 
supergravities.

It should be noted that our analysis is different
from that of \cite{BGL}. The arguments of the authors of ref.\cite{BGL}
go as follows. ``From the expression for the 11D supergravity plane 
wave solution $ds^2= dx^+ dx^- + (Q/r^7) (dx^-)^2 +dx_i^2$, one sees that 
$x^-$, the 
putative compact ``null'' direction, is in fact spacelike for $r< \infty$. 
Thus it seems that the gravitational effects automatically provide 
with the almost light-likeness of the circle which is crucial in Seiberg's 
derivation 
of M-atrix theory. It is the reason why supergravity solution is valid
at short distances.'' As we understand,  the discussion of refs.\cite{pol} seems
to imply that the 
DLCQ quantization 
of M-theory in  
such a plane wave background cannot yield a theory of pure 
D0-branes (an ordinary matrix SUSY quantum mechanics) because the 
circle is only asymptotically null and negative
momentum modes do not decouple. In  our
analysis the validity of supergravity solution at short distances follows
from the matrix theory conjecture and the classical equivalence of
11-D supergravities on a light-like and a spatial circles, and therefore, 
 our analysis is different from that of ref.\cite{BGL}.

Let us give some examples \cite{bbpt,CT} which illustrate \eq{relation2}. 
The potential for 
graviton-graviton scattering with
no longitudinal momentum transfer
in 11-D supergravity compactified
on a light-like circle
 reads \cite{bbpt}
\be
H_{l.c.}=-{N_{(2)}\ov R}{\sqrt{1-h v^2}-1\ov h}\ , \ \ \ \ \ h={15  N_{(1)}\ov 2R^2 M_P^9 r^7} \ ,
\la{corrie}
\ee
where $N_{(1)}/ R$ and $N_{(2)}/ R$ are  the logitudinal momenta $p_-$  of the 
 source and probe gravitons, $v$--their relative transverse velocity and
$r$-- their transverse separation.
The corresponding expression for the graviton-graviton scattering 
potential in 11-D supergravity compactified on a spatial circle reads
\be
H_{s.l}=-{N_{(2)}\ov \td R}{\sqrt{1-(1+\td h)\td v^2}-1\ov 1+\td h}
\ ,
\ \ \ \ \ \td h=
{15  N_{(1)}\ov 2\td R^2 \td M_P^9 \td r^7} \ .
\ee
Using  the relations 
\be
\td v \td M_P=v M_P \ , \ \ \ \ \  \td r \td M_P=r M_P \ ,
\la{rel1}
\ee
which follow from \eq{para}, it is easy
to show that $H_{l.c.}=\lim_{\td R\rightarrow 0} H_{s.l.}$.

The next example is a non-trivial one.  
Let us consider membrane -- membrane interaction in 
$D=11$  supergravity with the compact  light-like direction. Let $p_-^{(1)}=N_{(1)}/R$, $p_-^{(2)}=N_{(2)}/R$ and $m_{(1)}$, $m_{(2)}$ be the longitudinal
momenta and masses of  the source and probe membranes, respectively. 
In the case of zero longitudinal momentum transfer the interaction potential of  two membranes moving with the relative
transverse velocity $v$ and separated by a distance $r$ reads \cite{CT}
\bea
H_{l.c.}&=&\LB {m_{(1)}\ov p_-^{(1)}}\RB^2 p_-^{(2)} W^{-1}\bigg(1-\sqrt{\LB
1-W\LB {v p_-^{(1)} \ov m_{(1)}}\RB^2 \RB [1+\LB 1- {m_{(2)} p_-^{(1)}\ov 
m_{(1)}p_-^{(2)}}\RB^2 W]}\bigg) \non
&&+{p_-^{(2)}\ov 2} \LB {m_{(1)}\ov p_-^{(1)}}\RB^2 - {m_{(1)}m_{(2)}\ov
p_-^{(1)}} \ , \ \ \ \ \  W={3 m_{(1)}  \ov 2 p_-^{(1)} R M_P^6 r^5} \ .
\eea
The corresponding potential in the
spatially compactified $D=11$  supergravity reads
\footnote{In ref.\cite{CT} an 
equivalent potential was obtained from ten dimensional IIA supergravity. It
is easier to obtain it directly from 11-D supergravity, starting with the
membrane probe action $S=\int L dt=-\T_2 \int d^3x (\sqrt{-\det g_{mn}}
-C_{t  x_1 x_2 }-\dot x_{11} C_{x_{11} x_1 x_2})$ 
and performing Legendre transformation $L\rightarrow L'=L({\dot x_{11}}(p_{11}))
-{\dot x_{11}}p_{11}$ .}
\bea
H_{s.l.} 
&=&p_{11}^{(2)}-\td m_{(2)} \K^{-1} \sqrt{\K\LB {1\ov \W\sinh^2\td\beta +\K}-\td v^2\RB
\LB 1+ {(\K p_{11}'/\td m_{(2)})^2\ov \W\sinh^2\td\beta +\K}\RB} \non
&&+\td m_{(2)}(\K^{-1}-1)\cosh\td\beta - {\W  p_{11}'\cosh\td\beta \sinh\td\beta
\ov \W\sinh^2\td\beta +\K} \ ,
\eea
where 
\bea
&& 
p_{11}^{(1)}={N_{(1)}\ov \td R}=\td m_{(1)} \sinh\td\beta \ , \ \
p_{11}^{(2)}={N_{(2)}\ov \td R}\ , \ \ \ 
\td W={3\ov 2 \td R \td M_P^6 \cosh\td\beta \td r^5} \ , \non
&& \ \ \ \K =1+\td W \ , \ \ \ \ p_{11}'=p_{11}^{(2)}+ \td m_{(2)} (\K^{-1}-1)\sinh\td\beta \ .
\la{rel}
\eea
Using \eq{para}, \eq{rel1} and \eq{rel}, it can be  shown that
\bea 
&&\lim_{\td R\rightarrow 0}H_{s.l.} \non
&&= \lim_{\td R\rightarrow 0} p_{11}^{(2)} (\W \sinh^2\td\beta)^{-1} 
\bigg( 1-\sqrt {(1-\W \sinh^2\td\beta \td v^2) [1+\W\sinh^2\td\beta \LB 
{\td m_{(2)}\ov p_{11}^{(2)}}-{1\ov \sinh\td\beta}\RB^2 ]}\bigg) \non 
&&\ \ + \ \ {p_{11}^{(2)}\ov 2\sinh^2
\td\beta }
-{\td m_{(2)}\ov \sinh\td\beta}\non
&&=H_{l.c.} \ .
\eea
It would be interesting to extend these scattering potential calculations 
to  the cases 
involving longitudinal momentum exchange, recoil effects, etc. In view of
 relation \eq{relation2}, one expects agreement in these cases. 
Note that the equivalence between two supergravities holds for arbitrary $N$,
but in  relating M-atrix theory and SUGRA/SYM duality we 
use only the equivalence for
large N.\footnote{11-D supergravity on a light-like circle at finite-N 
is not necessarily the same as the low energy limit of DLCQ M-theory 
\cite{hell} and therefore the  equivalence of two supergravities at
finite-N is useless for our purposes.}

Let us conclude by summarizing our arguments. M-atrix theory conjecture
implies the equivalence of strongly coupled one dimensional SYM and
11-D light-cone supergravity. On the other hand, SUGRA/SYM duality
implies the equivalence of one dimensional SYM and 11-D 
supergravity on a spatial circle in Sen-Seiberg's limit. We showed 
that the hamiltonian of 11-D supergravity on a light-like circle is
equal to a limit of  the hamiltonian of 11-D supergravity on a spatial 
circle. In this way we argued that the 
large-N
M-atrix theory in the supergravity regime and  SUGRA/(0+1)-D SYM duality
are equivalent.

%%%%%%%%%%%%%%%%%%%%%%%%%%%%%%%%%%%%%%%%
%%%%%%%%%%%%%%%%%%%%%%%%%%%%%%%%%%%%%%%%%%%%%%%%%%%
%%%%%%%%%%%%%%%%%%%%%%%%%%%%%%%%%%%%%%%%%%%%%%%
\bigskip
\centerline {\ \bf Acknowledgments}
We are grateful to A.A.~Tseytlin for useful discussions. We also thank
J.~Polchinski and S-J.~Rey for the correspondence.
This work  was supported in part  by NSF  grant PHY-9309888. 

%%%%%%%%%%%%%%%%%%%%%%%%%%%%%%%%%%%%%%%%%%%%%%%%%%%%%%%%%%%

\bigskip

\setcounter{section}{0}
\setcounter{subsection}{0}

%%%%%%%%%%%%%%%%%%%%%%%%%%%%%%%%%%%%%%%%%%%%%%%%%%%%%%%%%%%%%%%%%%%%%%5


\begin{thebibliography}{9}




\bibitem{bfss}
 T. Banks, W. Fischler, S.H. Shenker and L. Susskind,
\pr D55 (1997) 5112, hep-th/9610043.

\bi{suss}
L. Susskind, {\it Another Conjecture about 
M(atrix)
Theory}, hep-th/9704080.

\bi{malda} J. Maldacena, {\it The large $N$ limit of superconformal field theories and supergravity}, 
Adv.Theor.Math.Phys. 2 (1998) 231-252,
hep-th/9711200.

\bi{doug}
M.R.~Douglas,
{\it D-branes and Matrix Theory in Curved Space},
Nucl.Phys.Proc.Suppl. 68 (1998) 381-393,
hep-th/9707228.


\bi{BGL}
V.~Balasubramanian, R.~Gopakumar and F.~Larsen,
{\it Gauge Theory, Geometry and the Large N Limit},
Nucl.Phys. B526 (1998) 415-431,
hep-th/9712077.


\bi{hyun}
S.~Hyun,
{\it The Background Geometry of DLCQ Supergravity},
Phys.Lett. B441 (1998) 116-122,
hep-th/9802026.


\bi{lif}
G.~Lifschytz,
{\it DLCQ-M(atrix) Description of String Theory, and Supergravity},
Nucl.Phys. B534 (1998) 83-95,
hep-th/9803191. 



\bi{HK}
S.~Hyun and Y.~Kiem,
{\it Background geometry of DLCQ M theory on a p-torus and holography},
Phys.Rev. D59 (1999) 026003,
hep-th/9805136. 


\bi{SM}
E.~Martinec and V.~Sahakian, 
{\it Black Holes and the Super Yang-Mills diagram. II},
hep-th/9810224.

\bi{pol}
J.~Polchinski,
{\it Light-cone or not light-cone},
Talk given at IAS, Princeton, December 1998.

\bi{halpern}
M.~Claudson and M.~Halpern, \np B250 (1985) 689;
R.~Flume, Ann. of Phys. 164 (1985) 189;
M.~Baake, P.~Reinicke and V.~Rittenberg, J.Math.Phys. 16 (1985) 1070.

\bi{sen}
A.~Sen,
{\it $D0$ Branes on $T^n$ and Matrix Theory},
Adv.Theor.Math.Phys. 2 (1998) 51-59,
hep-th/9709220.
 
\bi{seiberg}
N.~Seiberg,
{\it Why is the Matrix Model Correct?},
Phys. Rev. Lett. 79 (1997) 3577,
hep-th/9710009.

\bi{sen2}
A.~Sen,
{\it An Introduction to Non-perturbative String Theory},
hep-th/9802051.

\bi{gkp}
S.S.~Gubser, I.R.~Klebanov and A.M.~Polyakov,
{\it Gauge Theory Correlators from Non-Critical String Theory},
Phys.Lett. B428 (1998) 105-114,
hep-th/9802109.


\bi{wit}
E.~Witten,
{\it Anti De Sitter Space And Holography},
Adv.Theor.Math.Phys. 2 (1998) 253-291,
hep-th/9802150. 



 
\bi{IMSY}
N.~Itzhaki, J.M.~Maldacena, J.~Sonnenschein and S.~Yankielowicz,
{\it Supergravity and The Large N Limit of Theories With Sixteen Supercharges},
Phys.Rev. D58 (1998) 046004,
hep-th/9802042.

\bi{domain} 
H.J.~Boonstra, K.~Skenderis and P.K.~Townsend
{\it The domain-wall/QFT correspondence},
JHEP 9901 (1999) 003, hep-th/9807137; P.K.~Townsend,
{\it The M(atrix) model/$adS_2$ correspondence},
hep-th/9903043. 


\bi{msw}
J.~McCarthy, L.~Susskind and  A.~Wilkins,
{\it Large N and the Dine-Rajaraman problem},
Phys.Lett. B437 (1998) 62-68,
hep-th/9806136.

\bi{banks}
T.~Banks,
{\it Matrix Theory},
Nucl.Phys.Proc.Suppl. 67 (1998) 180-224,
hep-th/9710231.

\bibitem{bbpt} 
K.~Becker, M.~Becker, J.~Polchinski and A.A.~Tseytlin, 
{\em Higher order graviton scattering in M(atrix) theory}, 
Phys. Rev. D56 (1997) 3174, 
hep-th/9706072.

\bibitem{CT}
I.~Chepelev and A.A.~Tseytlin,
{\it On membrane interaction in matrix theory},
Nucl.Phys. B524 (1998) 69-85,
hep-th/9801120. 




\bi{hell}
S. Hellerman and J. Polchinski, {\it  Compactification 
in the light-like limit}, 
hep-th/9711037. 


\end{thebibliography}
\end{document}